\documentclass[journal]{IEEEtran}
\usepackage{amsmath}

\usepackage{cite}
\usepackage{graphicx,amssymb}
\usepackage{amsmath,amsfonts,amssymb}

\usepackage[usenames]{color}
\usepackage{algorithm}
\usepackage{algorithmic}
\usepackage{epstopdf}
\usepackage{multirow}
\usepackage{stfloats}

\begin{document}
 
\title{On the Capacity of Joint Fading and Two-path Shadowing Channels}

\author{I.~Dey,~\IEEEmembership{Student Member,~IEEE}, G.~G.~Messier,~\IEEEmembership{Member,~IEEE},~and~S.~Magierowski,~\IEEEmembership{Member,~IEEE}
\thanks{Copyright (c) 2013 IEEE. Personal use of this material is permitted. However, permission to use this material for any other purposes must be obtained from the IEEE by sending a request to pubs-permissions@ieee.org.}
\thanks{I.~Dey and G.~G.~Messier are with Department of Electrical and Computer Engineering, University of Calgary, 2500 University Drive N.W. Calgary, Alberta, Canada T2N 1N4 (E-mails: deyi and gmessier@ucalgary.ca).}
\thanks{S.~Magierowski is with the Department of Electrical Engineering and Computer Science, Lassonde School of Engineering, York University, Lassonde Bldg. 1012B, 4700 Keele St Toronto, Ontario, Canada M3J 1P3 (E-mail: magiero@cse.yorku.ca).} %
}


\maketitle

\begin{abstract}

The ergodic and outage channel capacity of different optimal and suboptimal combinations of transmit power and modulation rate adaptation strategies over a Joint fading and Two-path Shadowing (JFTS) fading/shadowing channel is studied in this paper. Analytically tractable expressions for channel capacity are obtained assuming perfect channel side information at the receiver and / or the transmitter with negligible feedback delay. Further, the impacts of the JFTS parameters on the channel capacity achieved by these adaptive transmission techniques are determined.

\end{abstract}

\IEEEpeerreviewmaketitle


\section{Introduction}\label{S1}

The high density and considerable individual data rate requirements of modern indoor wireless users has made high capacity wireless communications a priority in indoor environments.  While the use of indoor pico-cells is expected to grow, this demand is primarily being served today by indoor wireless access points.  Therefore, it is essential to have an accurate picture of what high throughput wireless communications systems can achieve when implemented on densely deployed indoor access points.

This picture is provided by Shannon channel capacity.  With the introduction of capacity achieving coding schemes \cite{3}, Shannon capacity is now of both theoretical and practical interest. In case of wireless links, Shannon channel capacity characterizes the long-term achievable information rate and therefore is termed as the ergodic capacity \cite{1}.

In a fading environment, the Shannon bound can be achieved by adapting a variety of parameters relative to the channel quality, if perfect channel side information (CSI) is available at the receiver and/or the transmitter \cite{5}, \cite{6}, \cite{7}. Examples include Optimal Rate Adaptation (ORA) \cite{8}, which adapts modulation constellation size, and Optimal Power and Rate Adaptation (OPRA) \cite{7}, which adapts a combination of modulation rate and transmit power. The Shannon capacity can also be achieved only through optimal power control by using fading inversion to maintain a constant carrier signal-to-noise ratio (CSNR). This technique is known as Channel Inversion with Fixed Rate (CIFR) \cite{7}, \cite{8}. Another adaptive transmission technique referred to as Truncated Channel Inversion with Fixed Rate (TIFR) is introduced in \cite{4}, where the channel fading is compensated only when the received CSNR is above a certain cut-off fade depth. The constant information rate that can be achieved using TIFR with an outage probability under a certain threshold is referred to as outage capacity \cite{9}.

It is important to point out that ergodic Shannon capacity estimates are only as good as the channel model upon which they are based.  It is well known that indoor wireless links are affected by both small scale fading and shadowing effects.  A Shannon capacity estimate is meant to characterize throughput experienced on time scales beyond a few seconds must be based on channel models that take both large and small scale effects into account.

To date, composite channel models that combine large and small scale effects have been developed primarily for outdoor channels.  In the bulk of these models, like the Suzuki \cite{10} and Nakagami - log-normal \cite{11} composite channel models, the log-normal distribution is used to model shadowing. A more practical closed-form composite fading model is the $\mathcal{K}$-distribution \cite{12}, \cite{13}, where log-normal shadowing is approximated by Gamma shadowing. This is done because LMS and macro-cellular communication users are highly mobile in an outdoor environment transiting through several scattering clusters. As a result, a range of main waves arrive at the mobile, the strength of each of which can be drawn from the log-normal or the Gamma distribution.

These outdoor composite models do not accurately characterize the indoor wireless LAN link, primarily because the path between the access point and users are too short for shadowing to be characterized by a log-normal distribution.  A new composite channel model, called the Joint Fading and Two Path Shadowing (JFTS) model, is proposed in \cite{14}.  Based on an extensive channel measurement campaign, the JFTS model is shown to be a more accurate model for the indoor wireless LAN channel than any other composite channel model proposed to date.

The primary contribution of this paper is to derive analytically tractable expressions for JFTS ergodic capacity under different adaptive transmission schemes.  These expressions will provide new insight into the behavior of ergodic capacity for indoor WLAN systems due to the nature of the JFTS model. The JFTS distribution is a convolution of the Rician fading distribution and the two-wave with diffused power (TWDP) shadowing model. The Rician distribution can be expressed in terms of circular bivariate Gaussian random variable with potentially non-zero mean, while the TWDP \cite{15} distribution is the sum of two half-Ricians. Hence, the JFTS capacity expressions for adaptive modulation techniques do not approach a non-fading channel for high values of carrier signal-to-noise ratio (CSNR).  This is a fundamentally different behavior from capacity expressions based on conventional channel models. 

The JFTS capacity expressions also have added values compared to similar capacity work based on other channel models.  Unlike the $\mathcal{K}$-fading model, the JFTS model has been verified using a practical measurement campaign.  The JFTS channel also has a closed form PDF expression as opposed to the Suzuki or Nakagami-log-normal channel models.  The parameters of the JFTS distribution can also be varied to represent a wide variety of channel conditions like no-fading (infinitely high fading parameter), no-shadowing (infinitely high shadowing parameter), heavy fading (low fading parameter) or heavy shadowing (low shadowing parameter). Hence, the capacity expressions evaluated over the JFTS channel model will provide us with the achievable ergodic capacity measures over a large variety of practical channel conditions, without assuming that the propagation environment is complex Gaussian distributed.

The second contribution of this paper is to explore the relationship between the optimal cut-off CSNR and the average received CSNR for JFTS faded/shadowed links when adaptive transmission techniques are applied. Our numerical results show that in presence of heavy fading and shadowing, the cut-off CSNR remains significantly lower than 1 even at high received CSNR, as opposed to traditional fading models \cite{1}. A lower cut-off CSNR will result in lower achievable channel capacity over a JFTS channel in comparison to other composite channel models imparting same severity in fading and/or shadowing. These results will be used to analyze how the JFTS channel capacity behaves in a fundamentally different way than the other composite fading/shadowing models prevalent in literature. Our results will also demonstrate the effect of JFTS parameters on the optimal achievable rate (capacity) assuming perfect CSI to be available at the transmitter and/or the receiver.

The rest of the paper is organized as follows. In Section~\ref{S2}, we present the probability density function (PDF) of the received instantaneous CSNR over a JFTS communication channel. In Section~\ref{S3}, expressions for the channel capacity under different adaptive transmission policies are derived. Numerical results are presented in Section~\ref{S4} followed by some concluding remarks in Section~\ref{S5}. 

\section{Joint Fading and Two-path Shadowing Model}\label{S2}

In an indoor wireless LAN (WLAN) communication scenario representing an open concept office or laboratory layout, the PDF $f_A(\alpha)$ of the received signal envelope, $\alpha(t)$, can be given by \cite{16},
\begin{align} \label{eq1}
f_A (\alpha) =&~\sum_{i = 1}^4 \frac{b_i \alpha}{2 P_1 P_2} \sum_{h = 1}^m \mathcal{R}~e^{- K - S_h - \frac{\alpha_2}{2 P_2 r_h^2}} \nonumber\\
&\cdot~[e^{S_h \Delta \mathsf{M}_i} I_0 \big(2\alpha \sqrt{K S_h (1 - \Delta \mathsf{M}_i) / P_1 P_2}\big) \nonumber\\
&+~e^{- S_h \Delta \mathsf{M}_i} I_0 \big(2\alpha \sqrt{K S_h (1 + \Delta \mathsf{M}_i) / P_1 P_2}\big)
\end{align}
where $\mathsf{M}_i = \cos ((i - 1) \pi / 7)$, $I_0$ is the zeroth order modified Bessel function of the first kind, $m$ is the quadrature order (determining approximation accuracy) and $\mathcal{R} = \frac{w_h}{|r_h|}~e^{\frac{r_h^2(2P_1 - 1)}{2 P_1}}$. The parameter $K$ is the small scale fading parameter, $S_h$ is the shadowing parameter, $\Delta$ is the shape parameter of the shadowing distribution, $P_1$ and $P_2$ are the mean-squared voltages of the diffused and the shadowed components respectively. An order of the shadowing distribution, $i$, of 4 is used. This is done because, the 4th order TWDP distribution \cite{17} was found to offer the best fit of the extracted shadowing distribution of the measurement campaign in \cite{1}. In (\ref{eq1}), $b_i = a_i I_0(1)$, where $a_1 = \frac{751}{17280}$, $a_2 = \frac{3577}{17280}$, $a_3 = \frac{49}{640}$ and $a_4 = \frac{2989}{17280}$. The multiplier $w_h$ denotes the Gauss-Hermite quadrature weight factors which is tabulated in \cite{18} and is given by, $w_h = (2^{m - 1} m! \sqrt{\pi}) / (m^2 [H_{m - 1}(r_h)]^2)$, where $H_{m - 1}(.)$ is the Gauss-Hermite polynomial with roots $r_h$ for $h = 1, 2, \dotso, m$. For our analysis, we have chosen $m = 20$, as is done for parameter estimation of composite gamma - log-normal fading channels in \cite{19}. In this case, the mean-squared value of the joint faded and two-path shadowed envelope, $A$, can then be calculated using (\ref{eq1}) and the integral solution from \cite[eq.~6.643.2, p.~709]{20} as,
\begin{align} \label{eq3}
\Omega_A =&~\mathrm{E}\{A^2\} =~\sum_{i = 1}^4 \sum_{h = 1}^{20} \frac{b_i P_2 \mathcal{R} r_h^4}{P_1^2}~e^{- K - S_h} \nonumber\\
&\cdot~\bigg[e^{S_h \Delta \mathsf{M}_i - K S_h (1 - \Delta \mathsf{M}_i) \frac{r_h^2}{2 P_1}}\big(P_1 + K S_h r_h^2 (1 - \Delta \mathsf{M}_i) \big) \nonumber\\
+&~\big(P_1 + K S_h r_h^2 (1 + \Delta \mathsf{M}_i) \big)~e^{- S_h \Delta \mathsf{M}_i - K S_h (1 + \Delta \mathsf{M}_i) \frac{r_h^2}{2 P_1}} \bigg].
\end{align}

Let us denote the instantaneous received CSNR as $\gamma$ and the average received CSNR as $\overline{\gamma}$. The expression for the PDF of the instantaneous CSNR per symbol over a JFTS faded/shadowed channel has been derived in \cite{16} in terms of the JFTS parameters, $K$, $S_h$ and $\Delta$. Putting (\ref{eq3}) back in that expression for the PDF of $\gamma$, the final expression can be obtained in terms of $\overline{\gamma}$ and $\Omega_A$ as,
\begin{align} \label{eq4}
f_{\gamma}(\gamma) = \sum_{h = 1}^{20} \frac{\Omega_A}{2 \gamma P_2 r_h^2} \bigg[1~-~e^{-\frac{\Omega_A \gamma}{2 \overline{\gamma} P_2 r_h^2}}\bigg].
\end{align}
In the next section, we will be using (\ref{eq4}) to obtain expressions for the achievable ergodic and outage capacities of a JFTS fading/shadowing communication channel with different adaptive transmission techniques.

\section{Analysis of Channel Capacity}\label{S3}

Given an average transmit power constraint, the optimal cut-off CSNR level ($\gamma_0$) for any adaptive transmission technique must satisfy the relationship \cite{6}, $\int_{\gamma_0}^{+\infty} \Big(\frac{1}{\gamma_0} - \frac{1}{\gamma}\Big) f_{\gamma} (\gamma) \mathrm{d}\gamma = 1$. If the received instantaneous CSNR level $\gamma$ falls below $\gamma_0$, data transmission will be suspended. In order to find the relationship between $\gamma$ and $\gamma_0$ for adaptive transmission over a JFTS faded / shadowed channel, we need to solve two integrals, \cite[eq.~3.351.6, p.~340]{20}
\begin{align} \label{eq6}
\mathcal{I}_1 =~\int_{\gamma_0}^{+\infty} f_{\gamma} (\gamma)~\mathrm{d}\gamma = \int_{\gamma_0}^{+\infty} \frac{\mathfrak{B}}{\gamma}~\mathrm{d}\gamma - \int_{\gamma_0}^{+\infty} \frac{\mathfrak{B}}{\gamma}~e^{-\frac{\mathfrak{B}\gamma}{\overline{\gamma}}}~\mathrm{d}\gamma \nonumber\\ =~\mathfrak{B}~\text{Ei}\bigg(-\frac{\mathfrak{B}\gamma_0}{\overline{\gamma}}\bigg) - \mathfrak{B}~\text{log}(\gamma_0)~~~~~
\end{align}
and
\begin{align} \label{eq7}
\mathcal{I}_2 = \int_{\gamma_0}^{+\infty} \frac{1}{\gamma} f_{\gamma} (\gamma) \mathrm{d}\gamma = \int_{\gamma_0}^{+\infty} \frac{\mathfrak{B}}{\gamma^2} \mathrm{d}\gamma - \int_{\gamma_0}^{+\infty} \frac{\mathfrak{B}}{\gamma^2} e^{-\frac{\mathfrak{B}\gamma}{\overline{\gamma}}} \mathrm{d}\gamma \nonumber\\
= - \frac{\mathfrak{B}}{\gamma_0} e^{-\frac{\mathfrak{B}\gamma_0}{\overline{\gamma}}} - \frac{\mathfrak{B}^2}{\overline{\gamma}}~\text{Ei}\bigg(-\frac{\mathfrak{B}\gamma_0}{\overline{\gamma}}\bigg) - \frac{\mathfrak{B}}{\gamma_0}
\end{align} 
where $\text{Ei}(\cdot)$ is the exponential integral given by \cite{21} and $\mathfrak{B} = \sum_{h = 1}^{20} \frac{\Omega_A}{2 P_2 r_h^2}$. Now, putting the integral solutions obtained in (\ref{eq6}) and (\ref{eq7}), back in the above mentioned relationship, we can find the equation which the optimal cut-off CSNR should satisfy for adaptive transmission. Therefore, in case of a JFTS faded/shadowed channel, $\gamma_0$ should satisfy the following relationship,
\begin{align} \label{eq8}
\bigg(\frac{\mathfrak{B}}{\gamma_0} + \frac{\mathfrak{B}^2}{\overline{\gamma}}\bigg) \text{Ei}\bigg(-\frac{\mathfrak{B}\gamma_0}{\overline{\gamma}}\bigg) + \frac{\mathfrak{B}}{\gamma_0} \bigg(1 - \text{log}(\gamma_0) + e^{-\frac{\mathfrak{B}\gamma_0}{\overline{\gamma}}}\bigg) = 1.
\end{align} 

\subsection{Ergodic Capacity}\label{S3.1}

\subsubsection{Optimal Power and Rate Adaptation (OPRA)}\label{S3.1.1}

Assuming perfect CSI at the transmitter and the receiver, the ergodic channel capacity $\langle C \rangle_{\text{OPRA}}$ in bits/sec under an average transmit power constraint is given by, $\langle C \rangle_{\text{OPRA}} = B\int_{\gamma_0}^{+\infty} \text{log}_2 \Big(\frac{\gamma}{\gamma_0}\Big) f_{\gamma} (\gamma) \mathrm{d}\gamma$, where $B$ (Hz) is the channel bandwidth and $\gamma_0$ is the optimal cut-off CSNR. A water-filling algorithm is used for optimal power adaptation given by $S(\gamma) = \frac{1}{\gamma_0} - \frac{1}{\gamma}$ for all $\gamma \geq \gamma_0$. The optimal rate adaptation sends a rate of $\text{log}_2 (\gamma / \gamma_0)$ bits/sec for a fade level of $\gamma$. In order to find the final expression for channel capacity per unit bandwidth over a JFTS faded / shadowed channel ($\langle C/B \rangle_{\text{OPRA}}^{\text{JFTS}}$ [bits/sec/Hz]), we need to solve four sets of integrals in,
\begin{align} \label{eq10}
\bigg\langle \frac{C}{B} \bigg\rangle_{\text{OPRA}}^{\text{JFTS}} = \frac{1}{\text{log}(2)} \bigg[\underbrace{\int_{\gamma_0}^{+\infty} \text{log}(\gamma) \frac{\mathfrak{B}}{\gamma}\mathrm{d}\gamma}_{\mathcal{I}_3}~~~~~~~~~~~~~~~ \nonumber\\
- \underbrace{\int_{\gamma_0}^{+\infty} \text{log}(\gamma_0) \frac{\mathfrak{B}}{\gamma}\mathrm{d}\gamma}_{\mathcal{I}_4} -~\underbrace{\int_{\gamma_0}^{+\infty} \text{log}(\gamma) \frac{\mathfrak{B}}{\gamma}e^{-\frac{\mathfrak{B}\gamma}{\overline{\gamma}}}\mathrm{d}\gamma}_{\mathcal{I}_5} \nonumber\\
+ \underbrace{\int_{\gamma_0}^{+\infty} \text{log}(\gamma_0) \frac{\mathfrak{B}}{\gamma}e^{-\frac{\mathfrak{B}\gamma}{\overline{\gamma}}}\mathrm{d}\gamma}_{\mathcal{I}_6}\bigg].
\end{align}
The expression in (\ref{eq10}) can be obtained in a tractable form through the following steps of integral solutions and mathematical manipulations. Firstly we can express,
\begin{align} \label{eq11}
\mathcal{I}_3 - \mathcal{I}_4 = \frac{\mathfrak{B}}{2} \text{log}^2(\gamma_0).
\end{align}
Using the identities, $\text{Ei}(- x) = - \Gamma(0, x) - \text{log}(x) + \frac{1}{2}(\text{log}(- x) - \text{log}(-\frac{1}{x}))$ and $\text{log}(- x) = \text{log}(x) + \imath\pi$, valid for $x > 0$ \cite{21} and \cite[eq.~4.452.1, p.~573]{20}, and assuming that, $(\mathfrak{B}\gamma_0 / \overline{\gamma}) > 0$ and $(\overline{\gamma} / \mathfrak{B}\gamma_0) > 0$ and after some algebraic manipulations, we can express,
\begin{align} \label{eq12}
\mathcal{I}_6 - \mathcal{I}_5 =~\mathfrak{B}~\text{log} (\gamma_0) \text{log} \bigg(\frac{\mathfrak{B}\gamma_0}{\overline{\gamma}}\bigg) + \mathfrak{B}\mathcal{E}~\text{log} (\gamma_0)~~~~~~~ \nonumber\\
- \frac{\mathfrak{B}}{2} \text{log}^2(\gamma_0) - \frac{\mathfrak{B}^2 \gamma_0}{\overline{\gamma}}~_3F_3\bigg(1, 1, 1; 2, 2, 2; -\frac{\mathfrak{B}\gamma_0}{\overline{\gamma}}\bigg).
\end{align}
where $\mathcal{E}$ is the Euler-Mascheroni constant with a numerical value of $\mathcal{E} \approx 0.577216$. Finally, using (\ref{eq11}) and (\ref{eq12}), the expression in (\ref{eq10}) can be obtained as,
\begin{align} \label{eq13}
\bigg\langle \frac{C}{B} \bigg\rangle_{\text{OPRA}}^{\text{JFTS}} =~\frac{\mathfrak{B}~\text{log} (\gamma_0)}{\text{log}(2)} \bigg[\text{log} \bigg(\frac{\mathfrak{B}\gamma_0}{\overline{\gamma}}\bigg) + \mathcal{E} \bigg]~~~~~~~~~~~ \nonumber\\ 
- \frac{\mathfrak{B}^2 \gamma_0}{\overline{\gamma}~\text{log}(2)}~_3F_3\bigg(1, 1, 1; 2, 2, 2; -\frac{\mathfrak{B}\gamma_0}{\overline{\gamma}}\bigg)
\end{align}
where $_pF_q (\cdot)$ is the generalized confluent hyper-geometric function \cite{21} and $p, q$ are integers.\\

\subsubsection{Optimal Rate Adaptation (ORA)}\label{S3.1.2}

Assuming perfect CSI at the receiver only, the ergodic channel capacity $\langle C \rangle_{\text{ORA}}$ in bits/sec with constant power over any composite fading and shadowing channel is given by, $\langle C \rangle_{\text{ORA}} = B \int_0^{+ \infty} \text{log}_2 (1 + \gamma) f_{\gamma} (\gamma) \mathrm{d}\gamma$. It is shown in \cite{2} that $\langle C \rangle_{\text{OPRA}}$ becomes equal to $\langle C \rangle_{\text{ORA}}$ when the transmit power is kept constant for OPRA. Using the identity $\text{log}(1 + y) = \text{log}(y) - \sum_{n = 1}^{+ \infty} \frac{(- 1)^{n}}{n y^n}$ for $|y| > 1$, we can solve the integral in the above definition \cite[eq.~3.351.2, p.~340]{20}. Using (\ref{eq4}), the final expression for channel capacity per unit bandwidth with ORA transmission ($\langle C / B \rangle_{\text{ORA}}^{\text{JFTS}}$) over a JFTS faded / shadowed communication link can be written as,
\begin{align} \label{eq15}
\bigg\langle \frac{C}{B} \bigg\rangle_{\text{ORA}}^{\text{JFTS}} =~\frac{1}{\text{log}(2)}\bigg[\int_0^{\infty} \sum_{n = 1}^{+ \infty} \frac{(- 1)^{n}}{n} \frac{\mathfrak{B}}{\gamma^{n + 1}} e^{-\frac{\mathfrak{B}\gamma}{\overline{\gamma}}}\mathrm{d}\gamma\bigg] \nonumber\\
= \sum_{n = 1}^{+ \infty} \frac{\mathfrak{B} \Gamma(- n)}{n~\text{log}(2)} \bigg(-\frac{\mathfrak{B}}{\overline{\gamma}}\bigg)^n.
\end{align}

It is evident from (\ref{eq13}) and (\ref{eq15}), that ergodic capacity over a JFTS distributed link depends on the mean-squared value of the joint faded and two-path shadowed envelope, $\Omega_{A}$. Now from (\ref{eq3}), we observe that $\Omega_{A}$ decreases exponentially with the increase either in $K$ or $S_h$ or both. In (\ref{eq13}), the capacity term is directly proportional to $\big[\text{log} \big(\frac{\mathfrak{B}\gamma_0}{\overline{\gamma}}\big) + \mathcal{E} \big]$. Hence, as $\Omega_{A}$ decreases, $\big|\text{log} \big(\frac{\mathfrak{B}\gamma_0}{\overline{\gamma}}\big)\big|$ increases, since $\Omega_{A} < 1$. As a result, the term $\big[\text{log} \big(\frac{\mathfrak{B}\gamma_0}{\overline{\gamma}}\big) + \mathcal{E} \big]$ increases with the increase in the fading and/or the shadowing parameters resulting in the overall increase in the ergodic capacity. Similar intuitive conclusions can also be made from (\ref{eq15}), where capacity increases with the decrease in $\Omega_{A}$, since $\big\langle \frac{C}{B} \big\rangle \propto \big(-\frac{\mathfrak{B}}{\overline{\gamma}}\big)^n$ for $n > 0$.\\

\subsubsection{Channel Inversion with Fixed Rate (CIFR)}\label{S3.1.3}

Assuming perfect CSI at the transmitter and the receiver, the channel capacity of this technique for any fading/shadowing communication link is given by, $\langle C \rangle_{\text{CIFR}} = B~\text{log}_2 \Big(1 + \frac{1}{\int_{0}^{+ \infty} \frac{1}{\gamma} f_{\gamma}(\gamma) \mathrm{d}\gamma} \Big)$. Using the integral solution from (\ref{eq7}), it can be shown that CIFR channel capacity is equal to zero for the JFTS channel. In that case, a large amount of transmitted power will be required to compensate for the deep channel fades if this technique is used for adaptive transmission. A better approach will be to use truncated channel inversion with fixed rate, the channel capacity for which has been derived in the next subsection.

\subsection{Outage Capacity}\label{S3.2}

\subsubsection{Truncated Channel Inversion with Fixed Rate (TIFR)}\label{S3.2.1}

In case of TIFR, channel fading is inverted only if the received instantaneous CSNR level is above the cut-off fade depth ($\gamma_0$). The channel capacity with TIFR over any fading channel is obtained by maximizing the outage capacity ($C_{\text{out}}$) over all possible $\gamma_0$ and can be expressed as, $C_{\text{TIFR}} = \text{max}_{\gamma_0} C_{\text{out}}$, where $C_{\text{out}}$ is the outage capacity. The outage channel capacity for a fading/shadowing channel can be calculated as, $\langle C_{\text{out}} \rangle_{\text{TIFR}} = B~\text{log}_2 \Big (1 + \frac{1}{\int_{\gamma_0}^{+ \infty} \frac{1}{\gamma} f_{\gamma}(\gamma) \mathrm{d}\gamma} \Big) (1 - P_{\text{out}})$, where $P_{\text{out}}$ is the outage probability. For a JFTS fading/shadowing channel, $P_{\text{out}}$ can be calculated as,
\begin{align} \label{eq18}
P_{\text{out}} =~\int_0^{\gamma_0} f_{\gamma} (\gamma)~\mathrm{d}\gamma = \mathfrak{B}~\text{log}(\gamma_0) - \mathfrak{B}~\text{Ei}\bigg(-\frac{\mathfrak{B}\gamma_0}{\overline{\gamma}}\bigg)
\end{align}
using the integral solution provided in (\ref{eq6}). Using (\ref{eq7}), we can evaluate the channel capacity with TIFR in a JFTS faded / shadowed communication link which can be expressed as,
\begin{align} \label{eq19}
\bigg\langle \frac{C_{\text{out}}}{B} \bigg\rangle_{\text{TIFR}}^{\text{JFTS}}~=~\bigg(1 + \mathfrak{B}~\text{Ei}\bigg(-\frac{\mathfrak{B}\gamma_0}{\overline{\gamma}}\bigg) - \mathfrak{B}~\text{log}(\gamma_0)\bigg) \nonumber\\ 
\cdot~\text{log}_2 \Bigg(1 -  \frac{\gamma_0 \overline{\gamma}}{\mathfrak{B} \overline{\gamma}~e^{-\frac{\mathfrak{B}\gamma_0}{\overline{\gamma}}} + \mathfrak{B}^2 \gamma_0~\text{Ei}\big(-\frac{\mathfrak{B}\gamma_0}{\overline{\gamma}}\big) + \mathfrak{B}\overline{\gamma}}\Bigg).
\end{align}

\section{Numerical Results and Discussion}\label{S4}

\begin{figure}[tp!]
\begin{center}
 \includegraphics[width=1.95\linewidth]{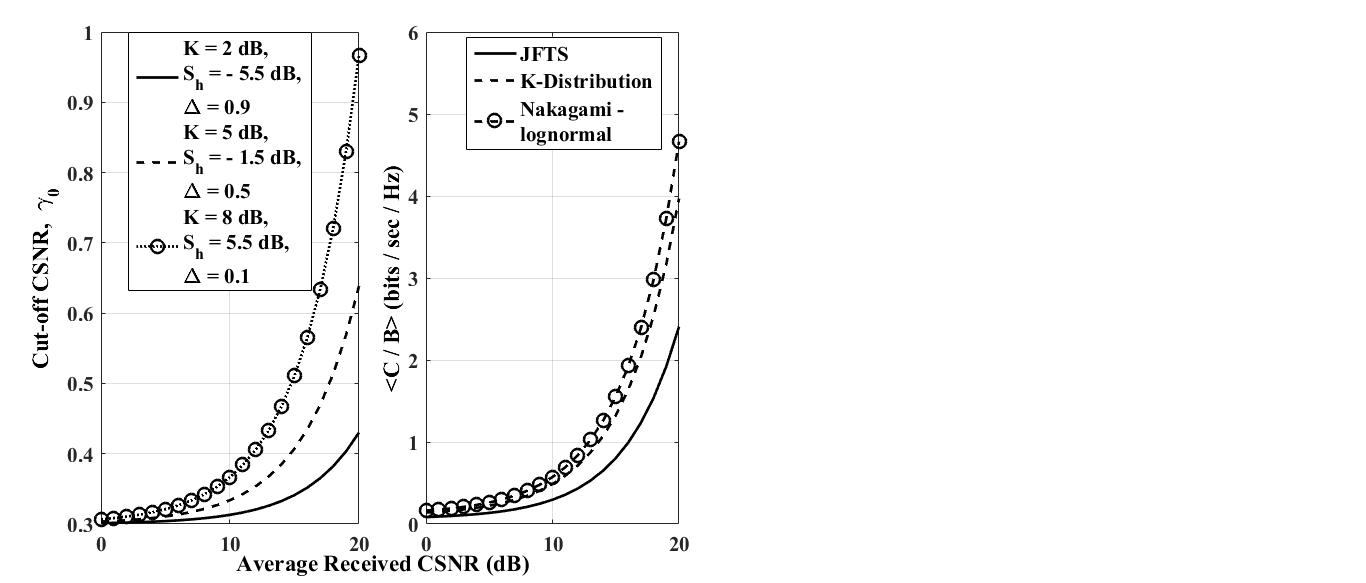}
\end{center}
\vspace*{-5mm}
\caption{(a) Calculated cut-off CSNR ($\gamma_0$) for different values of average received CSNR ($\overline{\gamma}$). (b) Ergodic Capacity per unit bandwidth achievable with OPRA over different channel models contributing the same AF of 3.45.}
\label{FIG1}
\vspace*{-2mm}
\end{figure}

It has been claimed in \cite{1} that for any fading channel, the optimal cut-off CSNR or optimal threshold satisfies $0 \leq \gamma_0 \leq 1$ if both the transmit power and the modulation rate are varied for optimal adaptation. Results from \cite{2, 3} also indicate that for Rayleigh and Nakagami-$m$ fading channels, $\gamma_0$ converges to 1 as $\overline{\gamma}$ increases. For a JFTS fading/shadowing channel, the relationship between $\gamma_0$ and $\overline{\gamma}$ is demonstrated in Fig.~\ref{FIG1}(a). For a communication link with high $K$ and $S_h$ (i.e. low fading and shadowing severity) $\gamma_0$ converges to 1 with the increase in $\overline{\gamma}$, as observed in \cite{1}. However, as the channel condition deteriorates with lower $K$ and $S_h$, $\gamma_0$ remains significantly lower than 1 even at high $\overline{\gamma}$. In such a scenario, perfect knowledge of both the transmit side and the receive side CSI should provide an edge over the perfect knowledge of only the receive side CSI, as claimed in \cite{1}. As a result, regulating both the transmit power and the modulation rate (OPRA) will result in a considerable increase in ergodic capacity over adapting only the modulation rate (ORA). This will be verified below.

The next set of curves in Fig.~\ref{FIG1}(b) are generated by comparing optimal achievable rate using OPRA over a JFTS channel with that achievable over conventional joint fading/shadowing channels like Nakagami-log-normal and $\mathcal{K}$-fading models. For each channel model, the distribution parameters are chosen such that the same amount of fading (AF) is contributed by each channel model. The curves are plotted for Nakagami-log-normal ($m = 1$, $\sigma = 3.88$) \cite{22}, $\mathcal{K}$-fading ($k = 0.96$) \cite{12}, and JFTS ($K = 5$ dB, $S_h = - 9.8$ dB, $\Delta = 0.1$) \cite{23} channel models, each contributing an AF of 3.45. An optimal cut-off CSNR, $\gamma_0$, which is significantly lower than 1 even at high $\overline{\gamma}$, results in lower achievable rate over a JFTS channel in comparison to other channel models. The reason can be attributed to the fact that the JFTS distribution has a very different PDF from common composite fading/shadowing distributions like Nakagami-log-normal or $\mathcal{K}$-distribution. Both of these distributions can be described using Gamma distribution and therefore the received envelope can be expressed in terms of zero mean complex Gaussian random variables with different shape factors. Hence in all of these cases, at higher received CSNR, the channel approaches the no-fading and no-shadowing condition and the received signal envelope becomes zero mean complex Gaussian distributed with a shape factor of 1. As a result, the achievable ergodic channel capacity starts approaching the Shannon bound as the received CSNR increases. While JFTS distribution can only be expressed in terms of bi-variate non-centralized chi-squared distribution and therefore can never be described using Gaussian random variables.

\begin{figure}[tp!]
\begin{center}
 \includegraphics[width=1.85\linewidth]{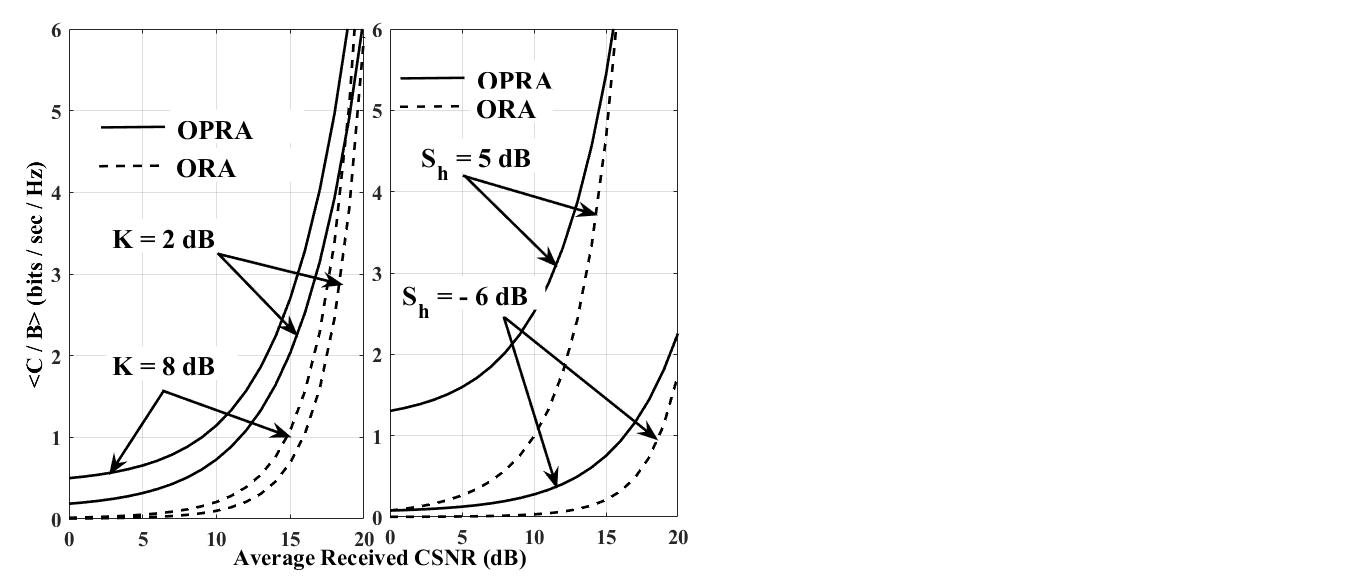}
\end{center}
\vspace*{-5mm}
\caption{Ergodic capacity per unit bandwidth of JFTS communication link with OPRA and ORA, where the curves are generated by (a) varying the $K$-factor ($S_h = - 2$ dB and $\Delta = 0.4$) and (b) varying the $S_h$-factor ($K = 5$ dB and $\Delta = 0.9$).}
\label{FIG2}
\vspace*{-2mm}
\end{figure}

It is claimed in \cite{2} that the difference in channel capacity between OPRA and ORA is bounded by $C_{\text{OPRA}} - C_{\text{ORA}} \geq B~\text{log}_2 \big(1 + \int_0^{\gamma_0} (\gamma - \gamma_0)f_{\gamma}(\gamma) \mathrm{d}\gamma\big)$. As a result, the channel capacity obtained using ORA starts approaching that achievable using OPRA with the increase in $\overline{\gamma}$ for JFTS channels, as is evident in  Fig.~\ref{FIG2}(a) and Fig.~\ref{FIG2}(b). Hence it can be concluded by summarizing the results from Fig.~\ref{FIG2}(a) and Fig.~\ref{FIG2}(b) that OPRA offers improvement in ergodic capacity over ORA only when $\gamma_0$ remains significantly lower than 1. These observations are similar to that made in \cite{2} and \cite{1} for Rayleigh and Nakagami-$m$ fading channels. The gap between $C_{\text{OPRA}}$ and $C_{\text{ORA}}$ increases at lower $\overline{\gamma}$ with the increase in severity of fading (decrease in $K$) and shadowing (decrease in $S_h$) both of which degrade the channel quality. These results are in accordance with the general behavior of a wireless communication system over a JFTS faded/shadowed channel. As noted in \cite{16}, performance of any communication system over a JFTS channel deteriorates with the decrease in $K$ and $S_h$-factors.

The degradation in ergodic capacity due to the decrease in $K$-factor from 8 dB to 2 dB (refer to Fig.~\ref{FIG2}(a)) is much less compared to the decrease in optimal achievable rate due to the lowering of $S_h$-factor from 5 dB to $- 6$ dB (refer to Fig.~\ref{FIG2}(b)). These results do not agree with the observations made in \cite{16}, where bit error rate performance of BPSK is found degrade equally either due to the decrease in the $K$-factor or the $S_h$-factor. The reason for this can be attributed to the $\Delta$-value chosen for each plot. For Fig.~\ref{FIG2}(a) a low $\Delta$ of 0.4 is chosen. In this case shadowing severity is reduced by the fact that only one scattering cluster dominates instead of two clusters. For Fig.~\ref{FIG2}(b) a high $\Delta$ of 0.9 is chosen, where the magnitudes of the shadowing values contributed by each scattering cluster are almost equal. As a result, even for a high  $\overline{\gamma}$ of 12 dB a penalty of 3 bits/sec/Hz of achievable rate is observed only for decreasing the $S_h$ factor.

\begin{figure}[tp!]
\begin{center}
\includegraphics[width=1.85\linewidth]{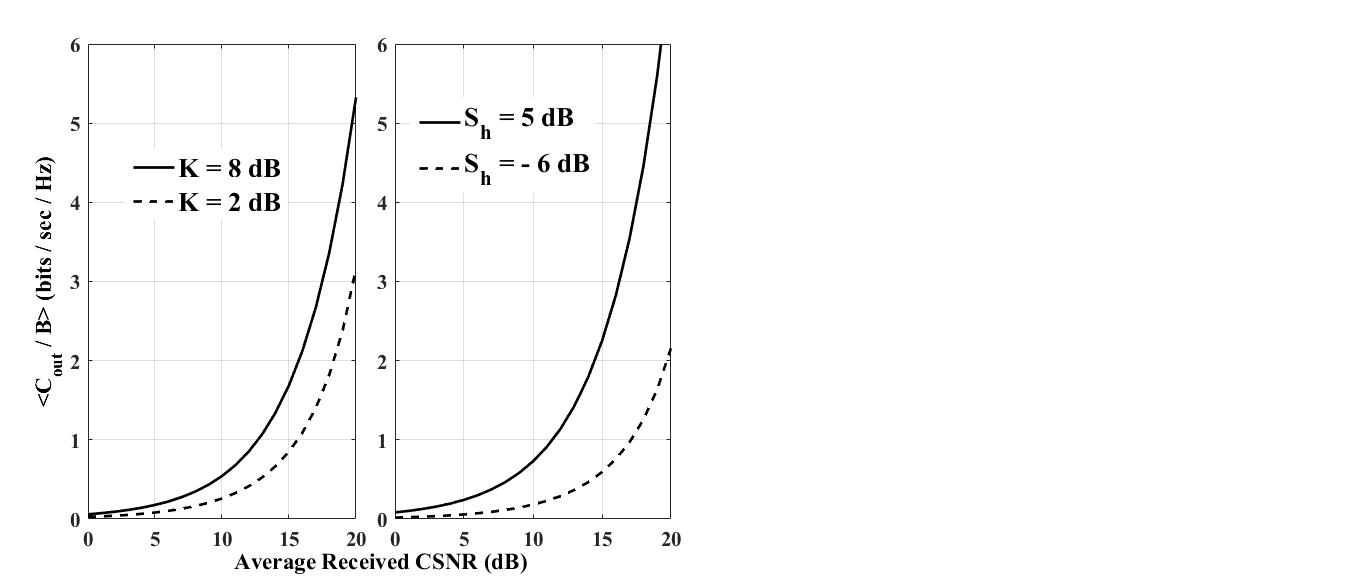}
\end{center}
\vspace*{-5mm}
\caption{Outage capacity per unit bandwidth of JFTS communication link with TIFR, where the curves are generated by (a) varying the $K$-factor ($S_h = - 2$ dB and $\Delta = 0.4$) and (b) varying the $S_h$-factor ($K = 5$ dB and $\Delta = 0.9$).}
\label{FIG3}
\vspace*{-2mm}
\end{figure}

On the other hand, the outage capacity with TIFR degrades equally with the lowering of either the small scale fading ($K$) factor or the shadowing ($S_h$) factor, as is evident in Fig.~\ref{FIG3}. Hence it can be concluded that the outage capacity of a JFTS communication channel is more sensitive than ergodic capacity to the changes in small scale fading and shadowing. This observation agrees with that made in case of Rician channel in presence of shadow fading in \cite{24}. It has also been observed in \cite{24}, increase in the severity of shadow fading improves ergodic capacity and degrades outage capacity of a shadowed Rician channel. However, for a JFTS faded/shadowed channel both ergodic and outage capacities are degraded significantly due to the increase in shadowing severity, as is evident from Fig.~\ref{FIG2} and Fig.~\ref{FIG3}.

\section{Conclusion}\label{S5}

The main aim of this paper is to derive the analytical expressions for achievable ergodic and outage channel capacities of different adaptive transmission techniques over JFTS fading/shadowing distribution assuming perfect CSI at the receiver and / or the transmitter. As a consequence, the effect of the JFTS parameters on the achievable channel capacities is also determined. Both ergodic and outage capacity decreases with a decrease in JFTS parameters $K$, $S_h$ and an increase in $\Delta$ while outage capacity is more sensitive than ergodic capacity to the changes in the JFTS parameters. Adaptation techniques like OPRA and ORA offer a considerable improvement in performance in comparison to CIFR and TIFR.

\end{document}